\documentclass[9pt,conference]{IEEEtran}
\IEEEoverridecommandlockouts

\usepackage{cite}
\usepackage{amsmath,amssymb,amsfonts}
\usepackage{algorithmic}
\usepackage{graphicx}
\usepackage{textcomp}
\usepackage[table]{xcolor}
\usepackage{fdsymbol}
\usepackage{multicol}
\usepackage{multirow}
\usepackage{caption}
\usepackage{url}
\usepackage[framemethod=TikZ]{mdframed}
\usepackage{booktabs}
\def\BibTeX{{\rm B\kern-.05em{\sc i\kern-.025em b}\kern-.08em
    T\kern-.1667em\lower.7ex\hbox{E}\kern-.125emX}}
\definecolor{revised}{RGB}{0, 0, 255}
\newcommand{\myccone}{\cellcolor[HTML]{f2f2f2}}

\begin{document}

\title{SILA: Signal-to-Language Augmentation for Enhanced Control in Text-to-Audio Generation}

\author{\IEEEauthorblockN{Sonal Kumar$^{\spadesuit\boxdot}$ \quad Prem Seetharaman$^{\boxdot}$ \quad Justin Salamon$^{\boxdot}$ \quad Dinesh Manocha$^{\spadesuit}$ \quad Oriol Nieto$^{\boxdot}$}
\IEEEauthorblockA{\textit{$^{\boxdot}$Adobe Research \quad $^{\spadesuit}$University of Maryland}\\
San Francisco, CA, USA - College Park, MD, USA \\
sonalkum@umd.edu\\
Project: \url{https://sonalkum.github.io/SILA/}
}
}

\maketitle

\begin{abstract}
The field of text-to-audio generation has seen significant advancements, and yet the ability to finely control the acoustic characteristics of generated audio remains under-explored. In this paper, we introduce a novel yet simple approach to generate sound effects with control over key acoustic parameters such as loudness, pitch, reverb, fade, brightness, noise and duration, enabling creative applications in sound design and content creation. These parameters extend beyond traditional Digital Signal Processing (DSP) techniques, incorporating learned representations that capture the subtleties of how sound characteristics can be shaped in context, enabling a richer and more nuanced control over the generated audio. Our approach is \textit{model-agnostic} and is based on learning the disentanglement between audio semantics and its acoustic features. Our approach not only enhances the versatility and expressiveness of text-to-audio generation but also opens new avenues for creative audio production and sound design. Our objective and subjective evaluation results demonstrate the effectiveness of our approach in producing high-quality, customizable audio outputs that align closely with user specifications. 
\end{abstract}


\section{Introduction}

Creating audio content, such as sound effects, music, or speech, customized to specific needs is vital for a wide range of applications, including augmented and virtual reality, game development, and video editing. Recent advancements in generative models have enabled high-quality audio generation, in both conditional~\cite{evans2024stableaudioopen,liu2023audioldm} or unconditional~\cite{ho2020denoising} settings. Specifically, conditional audio generation, which enables users to guide the process using other modalities such as language, offers greater flexibility for customized audio outputs. Natural language stands out as a particularly versatile means of conditioning audio generation, allowing users to specify detailed characteristics such as pitch, acoustic environment, and temporal structure. This task, known as text-to-audio (TTA) generation, focuses on producing audio based on natural language descriptions.

Recently, TTA models have achieved remarkable performance across various sub-domains such as speech~\cite{wang2023neural}, music~\cite{copet2024simple}, and sound effects~\cite{evans2024stableaudioopen}. However, fine-grained control over elements like loudness, reverb, pitch, etc beyond the core semantic content (e.g. ``explosion'' vs ``loud explosion'', ``coin cling'' vs ``reverberant coin cling'')  is an open problem, possibly due to limitations in the text encoder~\cite{ghosh2024compa}, lack of training data~\cite{ghosh2024compa} or the generative model itself~\cite{oh2023demand}. Professionals must be able to manipulate granular aspects of audio to achieve their creative vision. This lack of control over acoustic characteristics makes text-to-audio generation hard to incorporate into the workflows of professional sound artists and designers. 

{\noindent \textbf{Main Contributions.}} In this paper, we propose a simple, scalable and model-agnostic approach to enable fine-grained control via text prompts for sound effect generation. Our approach augments text prompts for training and inference with information extracted from the acoustic properties from audio. 
By incorporating detailed text descriptors of the audio’s acoustic characteristics into the captions, our method enhances the model’s ability to learn disentangled representations of these properties. This leads to improved fine-grained control over audio generation to control subtleties in \textit{loudness}, \textit{reverb}, \textit{pitch}, \textit{noise} and \textit{duration}. 
Thus, our proposed method is \emph{scalable} as it works with any audio data, and \emph{model-agnostic} since it is applicable to any generation-based model architecture which can be conditioned on text.
Moreover, our technique achieves better $CLAP_{scr}$ alignment~\cite{wu2024largescalecontrastivelanguageaudiopretraining} and a comparable $FAD$ score~\cite{kilgour2019frechetaudiodistancemetric} to other baselines, and achieves the highest score on subjective evaluation. 

\section{Related Work}
\label{sec:related_work}
{\noindent \textbf{Audio Generation.}} Recent advancements in audio generation predominantly utilize deep generative models, such as Generative Adversarial Networks (GANs)~\cite{Goodfellow2014GenerativeAN,  wang2017tacotronendtoendspeechsynthesis} and Variational Autoencoders (VAEs)~\cite{kingma2022autoencodingvariationalbayes}, to synthesize high-quality audio signals. These models are typically trained on large-scale datasets and learn to produce audio by capturing complex temporal dependencies. The closest work to ours is ~\cite{evans2024stableaudioopen}, where the authors employ a Diffusion Probabilistic Model for text-to-audio generation, leveraging transformer-based architectures to model long-range dependencies in the audio sequence. Unlike our approach, which uses a novel text-conditioning mechanism to enhance fine-grained control over generated sound properties, their method primarily focuses on sound generation without the ability to control the acoustic properties. Other approaches have explored the use of autoregressive models~\cite{kreuk2023audiogentextuallyguidedaudio, liu2024autoregressivediffusiontransformertexttospeech}, and non-autoregressive models~\cite{evans2024stableaudioopen} to balance between generation quality and computational efficiency.
\vspace{1mm}
\begin{figure*}[t]
    \centering
    \includegraphics[width=\linewidth]{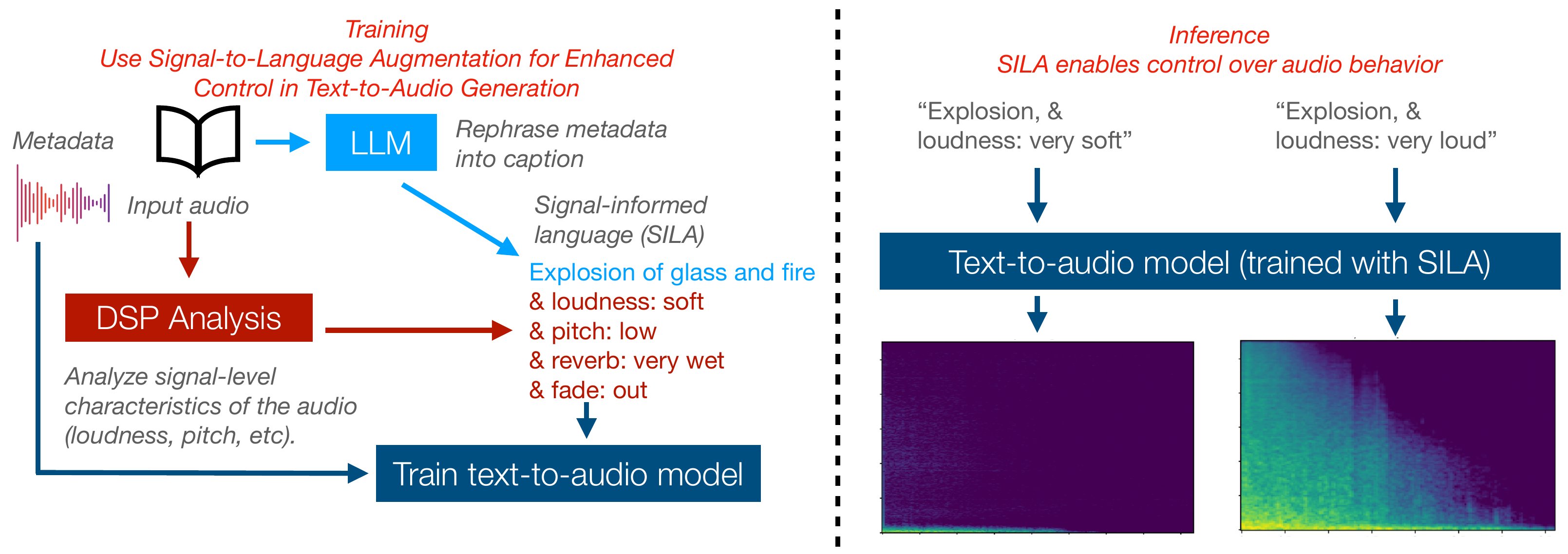}
    \caption{\small Illustration of our proposed methodology for training (left) and inference using (right) SILA. \textit{During training, the augmentation for reverb and fade is not always applied to the audio, and is only used for augmenting the audios to add \textit{reverb} and \textit{fade} descriptors.}}
    \label{fig:sila_fig}
\end{figure*}

{\noindent \textbf{Fine-Grained Control in Sound Generation.}} Specifically, this involves manipulating particular aspects of the audio output, such as pitch, timbre, loudness, and temporal structure, to create more expressive and customizable audio content. Recent advancements in this area leverage conditional generative models that incorporate auxiliary inputs or control signals to guide the generation process. For example, models like DiffWave~\cite{kong2020diffwave} and ~\cite{wu2024music} enable users to condition audio synthesis on various attributes, by incorporating control vectors or embedding layers that influence the generation at different stages. Other approaches use discrete and continuous latent representations \cite{hsu2018hierarchicalgenerativemodelingcontrollable, pmlr-v70-arik17a} to allow fine adjustments of the audio properties, enabling the generation of diverse and nuanced sounds from the same underlying model. These techniques often involve training with large, annotated datasets to learn the associations between control inputs and corresponding audio features. The ability to achieve fine-grained control not only enhances the flexibility and creativity of audio generation systems but also broadens their applicability in areas such as music production, virtual reality, and personalized content creation. As far as we are aware, ours is the first work to explore text-based fine-grained control in audio generation.

\section{Method}
\label{sec:method}

Fig.~\ref{fig:sila_fig} illustrates the methodology of our proposed method, SILA. In this section we describe SILA in detail, including its key components and training details.

\subsection{Caption Generation for SILA}

\label{subsec:caption_gen}
Our core idea is to generate captions with detailed descriptors about the acoustic characteristics of each event in the audio, such that when trained on such captions, the model learns to pay attention to these descriptors for generation. To achieve this, we first generate captions for each audio in the dataset. 

\noindent{\textbf{Coarse Caption Generation.}} We generate coarse captions for each audio in the dataset using GAMA \cite{ghosh2024gamalargeaudiolanguagemodel}, which is an audio language model that leverages Qformer based architecture to caption audio and perform complex reasoning on audio tasks. These preliminary captions are subsequently refined and expanded using the Mistral-7B model \cite{jiang2023mistral7b}, which is provided with additional metadata elements such as \textit{filename, title, description, keywords, category, subcategory, extras}, and the \textit{GAMA-generated caption}. Specifically, we utilize Mixtral-7B to generate multiple variations of a vivid and detailed description for each audio file. By using the carefully crafted prompt which utilizes the metadata, we ensure that the model captures diverse perspectives and rich, descriptive nuances of the audio, producing several distinct yet detailed interpretations of the sound. This approach allows us to explore the full expressive range of the generated descriptions while maintaining relevance to the audio’s acoustic characteristics. This approach is scalable and allows for the generation of captions for audios without ground-truth captions. Additionally, these captions are more diverse and accurate and better capture the nuances of the audio data.

\noindent{\textbf{DSP-Based Captions.}} Once coarse captions are generated, we concatenate audio descriptors about the acoustic characteristics of the audio at the end of the coarse caption obtained from the Mistral-7B model. These descriptors are appended to the caption as follows: 

\texttt{\textbf{\textit{Coarse caption, \& acoustic\_characteristics1: descriptor1, \& acoustic\_characteristics2: descriptor2, ....}}}

We discuss the audio descriptors used by SILA in the following subsection.

\subsection{Audio Descriptors}
\label{subsec:audio_desc}
Audio descriptors are measurable features that characterize various acoustic properties of audio signals. These descriptors serve a crucial role in understanding, analyzing, and manipulating audio, particularly in the context of text-to-audio generation. While traditional Digital Signal Processing (DSP) techniques have been extensively used to analyze and synthesize audio, our approach extends beyond the usage of simple DSP-based algorithms. Instead, we aim to learn and disentangle key acoustic attributes from the audio itself, allowing for more nuanced and flexible control over the generative process. These descriptors enable the model to capture and generate complex auditory characteristics, enhancing its ability to produce audio that aligns with specific user-defined properties. Below, we detail the descriptors used in our system.


\subsubsection{Loudness}
\label{subsec:loudness}
The loudness of each audio file is estimated and categorized into predefined classes based on an analysis of the training dataset. Following psychoacoustic principles outlined by Moore \cite{moore2012introduction}, loudness perception is influenced by sound pressure level and frequency. To this end, we define four distinct loudness categories: \textit{very soft} (-70 LKFS to -55 LKFS), \textit{soft} (-55 LKFS to -40 LKFS), \textit{loud} (-30 LKFS to -15 LKFS), and \textit{very loud} (greater than -15 LKFS). Audio files with intermediate loudness levels are not categorized to maintain focus on the most perceptually distinct variations. This classification approach enhances the model's ability to understand and differentiate between significant loudness levels, facilitating the generation of audio with accurate loudness descriptors.

\subsubsection{Pitch}
\label{subsec:pitch}
To accurately estimate the pitch, we employ a deep learning-based pitch estimation method, building on the work by Kim et al. \cite{kim2018crepeconvolutionalrepresentationpitch}, which utilizes convolutional neural networks for robust pitch detection across a wide range of frequencies. The audio is first processed through segmentation and normalization to ensure consistent pitch analysis. We then classify pitch into two categories: \textit{low} (less than 1.5 octaves) and \textit{high} (greater than 3.5 octaves), based on its frequency range. These classifications align with the frequency ranges associated with low and high pitch perception in human auditory studies, allowing the model to effectively understand and distinguish tonal variations in the audio. 

\subsubsection{Reverb}
\label{subsec:reverb}
Reverberation is an important auditory feature that adds depth and spatial characteristics to audio. Due to the lack of reverberant data in the training set, we augmented the dataset by applying simulated reverb effects to the audio using Pedalboard\footnote{https://github.com/spotify/pedalboard}, creating distinct categories such as \textit{dry}, \textit{slightly wet}, \textit{wet}, and \textit{very wet}. This augmentation strategy not only compensates for the absence of natural reverb in the training data but also enables the model to learn the nuances of different reverberation levels, thereby improving its ability to generate audio with realistic spatial characteristics.

\subsubsection{Noise}
\label{subsec:noise}
Noise estimation in audio is a challenging task due to the overlapping nature of background and foreground sounds. Traditional methods like root mean square (RMS) and zero-crossing rate (ZCR) often fail to differentiate between types of noise effectively. To address this, we developed a novel approach inspired by spectral analysis techniques. We compute the Mel spectrogram, normalize it, and calculate an estimate of SNR metric by taking the difference between the loudest set of frames and the softest set of frames to capture the dynamic range of the audio. We classify audio into \textit{silent background} (SNR $\geq$ 6) and \textit{noisy background} (SNR $\leq$ 2), providing a more nuanced representation of noise levels. This method allows the model to distinguish between different noise environments, aligning with findings in acoustic signal processing literature \cite{1163209}. We avoid augmenting the caption of audio with SNR estimates between $2$ and $6$, as we found this metric to be somewhat unreliable in the middle of the range. 

\subsubsection{Brightness}
\label{subsec:brightness}
Brightness, a perceptual attribute that reflects the presence of high-frequency content in audio, is estimated through spectral analysis. We calculate the spectral centroid, which indicates the center of mass of the spectrum, to determine the brightness level. Based on the distribution of spectral centroid values in the training data, audio files are classified into \textit{dull} or \textit{bright} categories. Audio with lower spectral centroid values (less than 45) are considered \textit{dull}, while those with higher values (greater than 65) are labeled as \textit{bright}. This classification helps the model distinguish between audio with varying amounts of high-frequency content, which is a key factor in the perception of brightness.

\subsubsection{Fade}
\label{subsec:fade}
Fade-in and fade-out effects, which involve gradual amplitude changes at the beginning or end of an audio signal, are common in audio production. To incorporate these effects, we augmented the audio data with synthesized fade-in and fade-out transitions. This augmentation allowed the model to learn to recognize and differentiate these common audio transitions, which are prevalent in various audio production contexts. By introducing these effects, the model gains the ability to generate audio that naturally incorporates fading, enhancing its applicability in real-world scenarios.

\subsubsection{Duration}
\label{subsec:duration}
Duration is an important contextual feature that provides information about the temporal length of an audio clip. To incorporate this information, the duration of each audio file was calculated and probabilistically appended to the audio's descriptive metadata. This addition allows the model to learn associations between audio content and its temporal length, improving its ability to generate audio that aligns with specific duration requirements. Including duration as a descriptor also enhances the model's understanding of time-dependent audio features, further refining its generative capabilities.

After adding these descriptors, a sample coarse caption \textit{"The deep rumble of the storm echoes through the sky."} looks like:
\textit{"The deep rumble of the storm echoes through the sky, \& loudness: soft, \& pitch: low, \& reverb: very wet, \& brightness: bright, \& fade: out, \& duration: 3 seconds."}

\subsection{Audio Generation}
\label{subsec:model_arch}
SILA relies on any audio generation model that can be conditioned on text, enabling fine-grained control over the generated audio without specifying a particular architecture. This flexibility allows the model to adapt to various generation tasks where text-based conditioning is essential, ensuring alignment between the descriptive prompts and the resulting acoustic outputs. The model’s conditioning mechanism captures key acoustic characteristics from the prompts, enabling precise control over features like loudness, pitch, reverb, and noise during the generation process. This modular approach ensures that the model can seamlessly integrate into different text-to-audio frameworks while providing robust and flexible control over the audio generation.

\section{Experiments}
\label{sec:exps}
{\noindent \textbf{Training Datasets.}} For training and evaluating SILA, we use FreeSound\cite{fonseca2022fsd50k} and licensed and proprietary sound effects data. We randomly split these datasets in train and validation. More detailed statistics about each dataset is provided on our project page.

{\noindent \textbf{Evaluation Datasets.}} We evaluate SILA on the Adobe Audition\footnote{https://www.adobe.com/products/audition/offers/audition-dlc.html} dataset.
It is an open-source dataset, and includes over 10,130 sound effects, categorized for various uses such as cartoon sounds, gunfire, crowd noises, and more, to enhance video, podcasting, or other audio projects. Note that AuditionSFX is not in our training data at all, and is used for evaluation purposes only.

\noindent{\textbf{Evaluation metrics.}} We evaluate the proposed method using both objective and subjective metrics. For the objective evaluation, we use two metrics: the Fréchet Audio Distance (FAD) and the CLAP score (CLAP). We report the FAD~\cite{kilgour2019frechetaudiodistancemetric} using the PyTorch implementation with the VGGish model, where a lower FAD score indicates more plausible generated audio. The CLAP score~\cite{wu2024largescalecontrastivelanguageaudiopretraining, huang2023makeanaudiotexttoaudiogenerationpromptenhanced} is computed between the track description and the generated audio to quantify audio-text alignment, using the official pretrained CLAP model. 

\noindent{\textbf{User-Evaluation.}} For the subjective evaluation, we surveyed 22 participants, all of whom were adults with no hearing impairment. Each participant was presented with 4 audios at once one from each model and was asked to choose one. In total the users were presented with 30 samples in random order. The participants had to choose which of the 4 audios aligned most to each of the categories: \textit{Loudness}, \textit{Pitch}, \textit{Reverb}, \textit{Noise}, \textit{Fade}, \textit{Duration}, and which of the 4 audios was overall most aligned to the caption used to generate the audio: \textit{All}. Importantly, the models used to generate the audios were unknown to the listeners, ensuring unbiased selections. This approach allowed us to evaluate both the specific acoustic control and the overall semantic alignment of the generated audio with the descriptive prompts.

{\noindent \textbf{Model Architecture and Hyper-parameters.}} Although SILA is model-agnostic, in our experiments we use our own DiT-based TTA model built on Diffusion Transformers (DiT)~\cite{peebles2023scalablediffusionmodelstransformers}.
This is a transformer-based architecture designed to operate on continuous data like audio, refining outputs through noise removal via diffusion processes. The DiT used is conditioned on language representations from FLAN-T5 XXL~\cite{chung2022scalinginstructionfinetunedlanguagemodels}, using cross-attention between T5 embeddings and the DiT representations at each layer. This enables a more precise alignment between textual descriptions and acoustic outputs, offering fine control over the generative process. Additionally, our model incorporates classifier-free guidance and dynamic conditioning mechanisms to further improve the quality, flexibility, and control of the generated audio. This enables our model to align textual descriptions with acoustic outputs in a highly flexible manner, allowing for fine control over the audio generation process. We train SILA with a learning rate of 1e-4 and an effective batch size (BS) of 64, and train for 400000 steps. We train a variational autoencoder (VAE) following the recipe in Kumar et al.~\cite{dac}, which compresses mono audio at 48khz to a latent space with 64 dimensions, running at 40Hz. Our only modification to the recipe in Kumar et al.~\cite{dac} is to swap the residual vector quantization with a continuous latent space that is KL-regularized.

{\noindent \textbf{Baselines.}} We use the following baselines for comparison: Stable Audio Open~\cite{evans2024stableaudioopen}, AudioGen~\cite{kreuk2023audiogentextuallyguidedaudio} and Tango 2~\cite{majumder2024tango2aligningdiffusionbased}. We compare our proposed approach to these baselines on Fréchet Audio Distance (FAD) and the CLAP score (CLAP). Due to computational constraints we use the checkpoints released by the authors.


\begin{table}[t]
\caption{\small Comparison of $CLAP_{scr}$ and $FAD$ scores on AuditionSFX. SILA achieves the highest $CLAP_{scr}$ and second highest $FAD$, indicating better alignment and plausible generation quality.}
\label{tab:results}
\centering
\begin{tabular}{lcc}
\toprule
Model       & $CLAP_{scr}\uparrow$ & $FAD\downarrow$ \\
\midrule \midrule
Stable Audio & {0.26} &  {0.86} \\
AudioGen     & \underline{0.27} &  {1.12}  \\
Tango 2       & 0.21 &  1.97 \\
Baseline (w/o SILA)  & 0.25 & \textbf{0.81} \\ 
\myccone SILA         &\myccone \textbf{0.29} &\myccone  \underline{0.84} \\\bottomrule
\end{tabular}
\end{table}
\section{RESULTS}
\label{sec:results}
Table~\ref{tab:results} compares the $CLAP_{scr}$ and $FAD$ scores across different models for the AuditionSFX dataset, showing that SILA achieves superior alignment between text descriptions and generated audio. The higher $CLAP_{scr}$ indicates that SILA is more effective in capturing and controlling the acoustic characteristics specified in the prompts compared to previous models. Importantly, this improvement in text-audio alignment is achieved without compromising the overall audio quality, as demonstrated by a competitive $FAD$ score, suggesting that SILA generates perceptually plausible audio while maintaining strong semantic alignment.

Table~\ref{tab:desc_res} presents a comparison of the scores between the baseline model (without SILA) and the model incorporating SILA. The results demonstrate that SILA enhances the model's ability to learn better disentanglement during training. This allows the model to capture diverse characteristics of the audio, beyond just the core content of the audio. We train both the model on the same datasets. For the comparison of \textit{noise}, \textit{brightness}, and \textit{pitch}, we use the metrics discussed in Section~\ref{subsec:audio_desc}. Since SILA adds a reverb effect to the audio, we calculate the reverb using timbral models\footnote{\url{https://github.com/AudioCommons/timbral_models}}, returning the mean RT60 value. Loudness is not included in the comparison as the model outputs normalized audio, and \textit{fade} values are omitted due to the absence of metrics for evaluating \textit{fade in} and \textit{fade out} effect.

In Table~\ref{tab:my-table}, the subjective evaluation further validates SILA's performance, as it outperforms other models in capturing key acoustic features such as \textit{loudness}, \textit{pitch}, \textit{reverb}, \textit{noise}, \textit{duration}, and \textit{fade}. Listeners consistently rated SILA's generated audio as aligning more closely with the descriptive prompts across all categories, highlighting its ability to disentangle and control individual audio features with greater precision. The results demonstrate SILA’s robustness in both objective and subjective metrics, confirming its effectiveness for practical applications where fine-grained control over audio characteristics is essential.

As demonstrated in Example 1 on the \textit{project website}, the model successfully disentangles loudness, producing soft explosion sounds that are perceptually soft rather than loud; other, subsequent examples on the project page strengthen our claim. It can also be observed that the soft explosion sound is not merely a quieter version of the loud one, but carries distinct auditory characteristics that make it perceptually \textit{different}. In a model trained with SILA, generated audio for a soft explosion is one that is in the distance, and far away, whereas a loud explosion is one that is very close.


\begin{table}[]
\caption{\small Comparison of average acoustic characteristic values. SILA outperforms the baseline, with scores within the expected range for each feature, indicating improved disentanglement between audio and its characteristics.}
\label{tab:desc_res}
\centering
\resizebox{1\columnwidth}{!}{
\begin{tabular}{lccc}
\toprule
Categories & Sub Categories    & Baseline (w/o SILA) & SILA \\
\midrule \midrule
\multirow{4}{*}{Reverb (RT60)}     & \textit{dry}  &     0.55                &   0.11   \\
           & \textit{slightly wet}      &  0.52    &   0.52   \\
           & \textit{wet}               &   0.50   &  0.73    \\
           & \textit{very wet}          &   0.51   &  0.84    \\ \midrule
\multirow{2}{*}{Noise (SNR)}      & \textit{silent background} &  4.31  &  6.78    \\
           & \textit{noisy background}  &       2.43  &  1.72    \\ \midrule
\multirow{2}{*}{Brightness (spec. centroid)} & \textit{dull}  &   57.53         &         41.71                 \\
           & \textit{bright}            &        59.24             &    71.37  \\ \midrule
\multirow{2}{*}{Pitch (octave)}   & \textit{low}         &     2.31     &   1.17   \\
           & \textit{high}         &    2.95    &   3.78   \\ \midrule
\multirow{2}{*}{Duration (sec.)}   & \textit{3 seconds}  &    4.87  &   3.12   \\
           & \textit{4 seconds}         &         4.94            &   3.97   \\ \bottomrule
\end{tabular}}
\end{table}

\begin{table}[]
\caption{\small Subjective evaluation comparing different baselines to SILA. Users show a preference for SILA over other models.}
\label{tab:my-table}
\centering
\resizebox{1\columnwidth}{!}{
\begin{tabular}{lccccccc}
\toprule
Model             & Loudness & Pitch & Reverb & Noise & Fade & Duration & All \\ \midrule \midrule
Stable Audio Open &    \underline{0.17}&   \underline{0.23}&   0.09     &  \underline{0.20}    &  0.18    &   \underline{0.26}       &  \underline{0.12}   \\
AudioGen          &    0.10&   0.17&   \underline{0.13}&  0.19    &  \underline{0.21}    &   0.22       &  0.11   \\
Tango 2           &    0.03      &   0.10    &   0.07     &  0.14    &  0.10    &    0.16      &  0.05   \\
\myccone SILA              &\myccone    \textbf{0.70}&\myccone  \textbf{0.50}&\myccone   \textbf{0.71}&\myccone  \textbf{0.47}    &\myccone \textbf{0.51}    &\myccone   \textbf{0.36}       &\myccone  \textbf{0.72}   \\ \bottomrule
\end{tabular}}
\end{table}

\section{Limitations and Future Work}
\label{sec:limitations}
We acknowledge the following limitations in our work:
\begin{enumerate}
    \item SILA has not been evaluated for compositional audio generation and may not perform well on audios containing multiple types of events.
    \item SILA has not been tested for controlling properties in text-to-speech generation.
\end{enumerate}
In future work, we aim to explore how SILA can be applied to control a wider range of audio properties such as stereo width, panning and motion (e.g. race car sound going from left to right) and to support compositional text-to-audio generation. We also hope to enable more natural captions where the characteristics are embedded in the caption (e.g. ``soft dog bark''), rather than appended to the end (e.g. ``dog bark \& loudness: soft'').

\section{Conclusion}
\label{sec:conclusion}
In this paper, we introduce SILA, a scalable and model-agnostic approach designed to enhance control over the acoustic characteristics of audio in the text-to-audio (TTA) generation process. Our method involves first generating diverse captions for the audio and subsequently incorporating audio descriptors that define the acoustic properties of the audio. This approach enables the TTA model to better learn the disentanglement between audio content and its acoustic features. Importantly, our method maintains the quality of audio generation while significantly improving the controllability of the generated audio.
\bibliography{custom}

\begin{thebibliography}{10}

\bibitem{evans2024stableaudioopen}
Zach Evans, Julian~D Parker, CJ~Carr, Zack Zukowski, Josiah Taylor, and Jordi Pons,
\newblock ``Stable audio open,''
\newblock {\em arXiv preprint arXiv:2407.14358}, 2024.

\bibitem{liu2023audioldm}
Haohe Liu, Zehua Chen, Yi~Yuan, Xinhao Mei, Xubo Liu, Danilo Mandic, Wenwu Wang, and Mark~D Plumbley,
\newblock ``{A}udio{LDM}: Text-to-audio generation with latent diffusion models,''
\newblock in {\em Proceedings of the 40th International Conference on Machine Learning}, Andreas Krause, Emma Brunskill, Kyunghyun Cho, Barbara Engelhardt, Sivan Sabato, and Jonathan Scarlett, Eds. 23--29 Jul 2023, vol. 202 of {\em Proceedings of Machine Learning Research}, pp. 21450--21474, PMLR.

\bibitem{ho2020denoising}
Jonathan Ho, Ajay Jain, and Pieter Abbeel,
\newblock ``Denoising diffusion probabilistic models,''
\newblock {\em Advances in neural information processing systems}, vol. 33, pp. 6840--6851, 2020.

\bibitem{wang2023neural}
Chengyi Wang, Sanyuan Chen, Yu~Wu, Ziqiang Zhang, Long Zhou, Shujie Liu, Zhuo Chen, Yanqing Liu, Huaming Wang, Jinyu Li, et~al.,
\newblock ``Neural codec language models are zero-shot text to speech synthesizers,''
\newblock {\em arXiv preprint arXiv:2301.02111}, 2023.

\bibitem{copet2024simple}
Jade Copet, Felix Kreuk, Itai Gat, Tal Remez, David Kant, Gabriel Synnaeve, Yossi Adi, and Alexandre D{\'e}fossez,
\newblock ``Simple and controllable music generation,''
\newblock {\em Advances in Neural Information Processing Systems}, vol. 36, 2024.

\bibitem{ghosh2024compa}
Sreyan Ghosh, Ashish Seth, Sonal Kumar, Utkarsh Tyagi, Chandra Kiran~Reddy Evuru, Ramaneswaran S, S~Sakshi, Oriol Nieto, Ramani Duraiswami, and Dinesh Manocha,
\newblock ``Compa: Addressing the gap in compositional reasoning in audio-language models,''
\newblock in {\em The Twelfth International Conference on Learning Representations}, 2024.

\bibitem{oh2023demand}
Sangshin Oh, Minsung Kang, Hyeongi Moon, Keunwoo Choi, and Ben~Sangbae Chon,
\newblock ``A demand-driven perspective on generative audio ai,''
\newblock {\em arXiv preprint arXiv:2307.04292}, 2023.

\bibitem{wu2024largescalecontrastivelanguageaudiopretraining}
Yusong Wu, Ke~Chen, Tianyu Zhang, Yuchen Hui, Taylor Berg-Kirkpatrick, and Shlomo Dubnov,
\newblock ``Large-scale contrastive language-audio pretraining with feature fusion and keyword-to-caption augmentation,''
\newblock in {\em ICASSP 2023-2023 IEEE International Conference on Acoustics, Speech and Signal Processing (ICASSP)}. IEEE, 2023, pp. 1--5.

\bibitem{kilgour2019frechetaudiodistancemetric}
Kevin Kilgour, Mauricio Zuluaga, Dominik Roblek, and Matthew Sharifi,
\newblock ``Fr$\backslash$'echet audio distance: A metric for evaluating music enhancement algorithms,''
\newblock {\em arXiv preprint arXiv:1812.08466}, 2018.

\bibitem{Goodfellow2014GenerativeAN}
Ian~J. Goodfellow, Jean Pouget-Abadie, Mehdi Mirza, Bing Xu, David Warde-Farley, Sherjil Ozair, Aaron~C. Courville, and Yoshua Bengio,
\newblock ``Generative adversarial nets,''
\newblock in {\em Neural Information Processing Systems}, 2014.

\bibitem{wang2017tacotronendtoendspeechsynthesis}
Yuxuan Wang, RJ~Skerry-Ryan, Daisy Stanton, Yonghui Wu, Ron~J Weiss, Navdeep Jaitly, Zongheng Yang, Ying Xiao, Zhifeng Chen, Samy Bengio, et~al.,
\newblock ``Tacotron: Towards end-to-end speech synthesis,''
\newblock {\em arXiv preprint arXiv:1703.10135}, 2017.

\bibitem{kingma2022autoencodingvariationalbayes}
Diederik~P Kingma,
\newblock ``Auto-encoding variational bayes,''
\newblock {\em arXiv preprint arXiv:1312.6114}, 2013.

\bibitem{kreuk2023audiogentextuallyguidedaudio}
Felix Kreuk, Gabriel Synnaeve, Adam Polyak, Uriel Singer, Alexandre D{\'e}fossez, Jade Copet, Devi Parikh, Yaniv Taigman, and Yossi Adi,
\newblock ``Audiogen: Textually guided audio generation,''
\newblock {\em arXiv preprint arXiv:2209.15352}, 2022.

\bibitem{liu2024autoregressivediffusiontransformertexttospeech}
Zhijun Liu, Shuai Wang, Sho Inoue, Qibing Bai, and Haizhou Li,
\newblock ``Autoregressive diffusion transformer for text-to-speech synthesis,''
\newblock {\em arXiv preprint arXiv:2406.05551}, 2024.

\bibitem{kong2020diffwave}
Zhifeng Kong, Wei Ping, Jiaji Huang, Kexin Zhao, and Bryan Catanzaro,
\newblock ``Diffwave: A versatile diffusion model for audio synthesis,''
\newblock in {\em International Conference on Learning Representations}, 2021.

\bibitem{wu2024music}
Shih-Lun Wu, Chris Donahue, Shinji Watanabe, and Nicholas~J Bryan,
\newblock ``Music controlnet: Multiple time-varying controls for music generation,''
\newblock {\em IEEE/ACM Transactions on Audio, Speech, and Language Processing}, vol. 32, pp. 2692--2703, 2024.

\bibitem{hsu2018hierarchicalgenerativemodelingcontrollable}
Wei-Ning Hsu, Yu~Zhang, Ron~J Weiss, Heiga Zen, Yonghui Wu, Yuxuan Wang, Yuan Cao, Ye~Jia, Zhifeng Chen, Jonathan Shen, et~al.,
\newblock ``Hierarchical generative modeling for controllable speech synthesis,''
\newblock {\em arXiv preprint arXiv:1810.07217}, 2018.

\bibitem{pmlr-v70-arik17a}
Sercan~{\"O}. Ar{\i}k, Mike Chrzanowski, Adam Coates, Gregory Diamos, Andrew Gibiansky, Yongguo Kang, Xian Li, John Miller, Andrew Ng, Jonathan Raiman, Shubho Sengupta, and Mohammad Shoeybi,
\newblock ``Deep voice: Real-time neural text-to-speech,''
\newblock in {\em Proceedings of the 34th International Conference on Machine Learning}, Doina Precup and Yee~Whye Teh, Eds. 06--11 Aug 2017, vol.~70 of {\em Proceedings of Machine Learning Research}, pp. 195--204, PMLR.

\bibitem{ghosh2024gamalargeaudiolanguagemodel}
Sreyan Ghosh, Sonal Kumar, Ashish Seth, Chandra Kiran~Reddy Evuru, Utkarsh Tyagi, S~Sakshi, Oriol Nieto, Ramani Duraiswami, and Dinesh Manocha,
\newblock ``{GAMA}: A large audio-language model with advanced audio understanding and complex reasoning abilities,''
\newblock in {\em Proceedings of the 2024 Conference on Empirical Methods in Natural Language Processing}, Yaser Al-Onaizan, Mohit Bansal, and Yun-Nung Chen, Eds., Miami, Florida, USA, Nov. 2024, pp. 6288--6313, Association for Computational Linguistics.

\bibitem{jiang2023mistral7b}
Albert~Q Jiang, Alexandre Sablayrolles, Arthur Mensch, Chris Bamford, Devendra~Singh Chaplot, Diego de~las Casas, Florian Bressand, Gianna Lengyel, Guillaume Lample, Lucile Saulnier, et~al.,
\newblock ``Mistral 7b,''
\newblock {\em arXiv preprint arXiv:2310.06825}, 2023.

\bibitem{moore2012introduction}
Brian~CJ Moore,
\newblock {\em An introduction to the psychology of hearing},
\newblock Brill, 2012.

\bibitem{kim2018crepeconvolutionalrepresentationpitch}
Jong~Wook Kim, Justin Salamon, Peter Li, and Juan~Pablo Bello,
\newblock ``Crepe: A convolutional representation for pitch estimation,''
\newblock in {\em 2018 IEEE International Conference on Acoustics, Speech and Signal Processing (ICASSP)}. IEEE, 2018, pp. 161--165.

\bibitem{1163209}
S.~Boll,
\newblock ``Suppression of acoustic noise in speech using spectral subtraction,''
\newblock {\em IEEE Transactions on Acoustics, Speech, and Signal Processing}, vol. 27, no. 2, pp. 113--120, 1979.

\bibitem{fonseca2022fsd50k}
Eduardo Fonseca, Xavier Favory, Jordi Pons, Frederic Font, and Xavier Serra,
\newblock ``Fsd50k: an open dataset of human-labeled sound events,''
\newblock {\em IEEE/ACM Transactions on Audio, Speech, and Language Processing}, vol. 30, pp. 829--852, 2021.

\bibitem{huang2023makeanaudiotexttoaudiogenerationpromptenhanced}
Rongjie Huang, Jiawei Huang, Dongchao Yang, Yi~Ren, Luping Liu, Mingze Li, Zhenhui Ye, Jinglin Liu, Xiang Yin, and Zhou Zhao,
\newblock ``Make-an-audio: Text-to-audio generation with prompt-enhanced diffusion models,''
\newblock in {\em International Conference on Machine Learning}. PMLR, 2023, pp. 13916--13932.

\bibitem{peebles2023scalablediffusionmodelstransformers}
William Peebles and Saining Xie,
\newblock ``Scalable diffusion models with transformers,''
\newblock in {\em Proceedings of the IEEE/CVF International Conference on Computer Vision}, 2023, pp. 4195--4205.

\bibitem{chung2022scalinginstructionfinetunedlanguagemodels}
Hyung~Won Chung, Le~Hou, Shayne Longpre, Barret Zoph, Yi~Tay, William Fedus, Yunxuan Li, Xuezhi Wang, Mostafa Dehghani, Siddhartha Brahma, et~al.,
\newblock ``Scaling instruction-finetuned language models,''
\newblock {\em Journal of Machine Learning Research}, vol. 25, no. 70, pp. 1--53, 2024.

\bibitem{dac}
Rithesh Kumar, Prem Seetharaman, Alejandro Luebs, Ishaan Kumar, and Kundan Kumar,
\newblock ``High-fidelity audio compression with improved rvqgan,''
\newblock in {\em Proceedings of the 37th International Conference on Neural Information Processing Systems}, Red Hook, NY, USA, 2024, NIPS '23, Curran Associates Inc.

\bibitem{majumder2024tango2aligningdiffusionbased}
Navonil Majumder, Chia-Yu Hung, Deepanway Ghosal, Wei-Ning Hsu, Rada Mihalcea, and Soujanya Poria,
\newblock ``Tango 2: Aligning diffusion-based text-to-audio generations through direct preference optimization,''
\newblock in {\em Proceedings of the 32nd ACM International Conference on Multimedia}, 2024, pp. 564--572.

\end{thebibliography}
\bibliographystyle{bib}
\end{document}